\documentclass[showpacs,preprintnumbers,amsmath,amssymb,letterpaper,cites]{revtex4}
\usepackage{graphicx}% Include figure files
\usepackage{enumerate}% For non-numerical items
\usepackage{bm}% bold math
\usepackage{slashed}
\usepackage{comment}

\newcommand\Eq[1]{Eq.~\ref{eq:#1}}

\usepackage{amssymb}
\usepackage{epsfig}
\newcommand{\be}{\begin{equation}}
\newcommand{\ee}{\end{equation}}
\newcommand\beq{\begin{eqnarray}}
\newcommand\eeq{\end{eqnarray}} 
\newcommand\eqn[1]{\label{eq:#1}} 
\newcommand\eq[1]{eq.~(\ref{eq:#1})} 
\newcommand\eqstwo[2]{eqs. (\ref{eq:#1},\ref{eq:#2})} 
\newcommand\eqs[2]{eqs. (\ref{eq:#1}-\ref{eq:#2})} 
\newcommand{\vev}[1]{\langle #1 \rangle}

\newcommand{\bfx}{{\mathbf x}}

\newcommand{\bfk}{{\mathbf k}}

\newcommand{\bfp}{{\mathbf p}}

\newcommand{\CA}{{\cal A}}

\newcommand{\CO}{{\cal O}}
\newcommand{\CM}{{\cal M}}
\newcommand{\CN}{{\cal N}}
\newcommand{\CP}{{\cal P}}

\newcommand{\CS}{{\cal S}}

\newcommand{\CL}{{\cal L}}

\newcommand{\Tr}{{\rm Tr\,}}
\newcommand\half{{\textstyle{\frac{1}{2}}}} 
 
\newcommand\expect[3]{\langle #1|#2|#3\rangle}

\newcommand{\mybar}[1]%
        {\kern 0.6pt\overline{\kern -0.6pt#1\kern -0.6pt}\kern 0.6pt}
\begin{document}

\preprint{INT-PUB-12-040}
\preprint{UM-DOE/ER/40762-525}

\title{Sign problems, noise, and chiral symmetry breaking in a QCD-like theory
}

\author{Dorota Grabowska$^1$}
\email{grabow@uw.edu}

\author{David B. Kaplan$^1$}
\email{dbkaplan@uw.edu}

\author{Amy N. Nicholson$^2$}
 \email{amynn@umd.edu}

\affiliation{$^1$Institute for Nuclear Theory, University of Washington, Seattle, Washington 98195-1550, USA}
 
\affiliation{$^2$ Maryland Center for Fundamental Physics, Department of Physics,
University of Maryland, College Park, Maryland 20742-4111, USA}


 %\date{\today}
 
 \begin{abstract}
The Nambu-Jona-Lasinio model reduced to 2+1 dimensions  has two different path integral formulations: at finite chemical potential one formulation has a severe sign problem similar to that found in QCD, while the other does not.   At large $N$, where $N$ is the number of flavors, one can compute the probability distributions of fermion correlators analytically in both formulations.  In the former case one finds a broad distribution with small mean; in the latter one finds a heavy tailed positive distribution amenable to the cumulant expansion techniques developed in earlier work.  We speculate on the implications of this model for QCD.  
 
\end{abstract}

\maketitle

\section{Introduction}
QCD has been around for 40 years and yet its application to the properties of bulk matter at nuclear density has proven to be stubbornly intractable.  An inherently nonperturbative problem, the only reliable tool available for solving QCD in this energy regime is Monte Carlo evaluation of lattice QCD. Unfortunately, lattice methods suffer from a severe sign problem in the grand canonical formulation that renders Monte Carlo techniques useless.  In a canonical formulation, despite recent impressive progress in studying light nuclei and hypernuclei \cite{Beane:2012vq}, there remain severe problems with signal-to-noise ratios; these problems are clearly related to the sign problem.  We continue here the research program outlined in Refs.~\cite{Lee:2011sm,Endres:2011jm,Endres:2011mm} where we study the probability distributions of correlators and argue that the origins of the sign or noise problem  lie in the dynamics and spectrum of the theory, and that it is not especially useful to think of it as a ``fermion" sign problem. In the particular case of QCD, the severity of the sign problem is closely related to the phenomenon of chiral symmetry breaking and the consequential existence of a light pion.  

In this paper we elucidate the connection between the QCD sign problem and chiral symmetry breaking by studying a simpler theory -- the Nambu-Jona-Lasinio (NJL) model \cite{Nambu:1961tp} in three dimensions. This formulation of the NJL model with $N$ flavors is of particular interest because it is soluble in large $N$, because it exhibits chiral symmetry breaking without the complication of confinement, and because it has two complementary but equivalent path integral formulations -- one of which looks very QCD-like and has a severe sign problem, while the other resembles more closely the chiral quark model \cite{Manohar:1983md} and has no sign problem.  

We begin by reviewing some features of the sign problem in QCD, and then turn to our analysis of the NJL model.  In particular we are able to analytically compute probability distributions for correlator measurements and show how the noise spectrum in these measurements is related to the presence or absence of a sign problem in the grand canonical formulation.  Our results should be directly applicable to suitable lattice formulations of this theory, but all of our analysis is analytic and in continuum Euclidean spacetime \footnote{We will identify $x_1$, $\partial_1$, and $\gamma_1$ with the Euclidean time direction; $\gamma$ matrices satisfy $\{\gamma_\mu,\gamma_\nu\}=2\delta_{\mu\nu}$.}.

At the end we return to QCD and speculate on the implications of our analysis for finding a path integral formulation of QCD that would allow for numerical study of bulk matter. 

\section{The sign and noise problems in QCD and the unique role of the pion}
 
Numerical simulations of lattice QCD  involve an approximation of the Feynman path integral in Euclidean spacetime. This requires a Monte Carlo evaluation of the averages of operators of interest - such as hadron correlation functions - over an ensemble of background gauge field variables generated with probability distribution
 \beq
\CP(U)\propto e^{-S[U]} \Delta[U]
 \eqn{prob}\eeq
 where $U$ is a link variable for the gluon degrees of freedom, $S$ is a suitable discretization of the Yang-Mills action, and $\Delta$ is the fermion determinant; both $S$ and $\Delta$ are functionals of the link variable $U$.  This program has been very successful for determining properties of the QCD spectrum in the vacuum, but it runs into an obstacle when trying to simulate matter at finite density in a grand canonical ensemble.  The fermion determinant $\Delta$ is a discretized version of $\det (\slashed{D} -m + \mu\gamma_1)$, where $D_\mu$ is the covariant derivative, $m$ is the quark mass, $\mu$ is the chemical potential for the quark number, and the gamma matrices $\gamma_1,\ldots,\gamma_4$ are all Hermitian.  At nonzero $\mu$ the fermion operator is in general complex, since $\mu\gamma_1$ is Hermitian while $\slashed{D}$ is anti-Hermitian, and the two do not commute. As a result the expression in \eq{prob} is not a suitable probability measure for the gauge field configurations; this problem has nothing to do with discretization of the lattice action, since it is a property of the continuum functional.  Therefore, for the rest of this paper we will only discuss the sign problem in the continuum, as the discretization of spacetime is not directly relevant (although a poor choice of discretization can in principle introduce additional sign problems not present in the continuum theory). 
 
At nonzero chemical potential one can write $\Delta(A_\mu,\mu) = |\Delta(A_\mu,\mu)|e^{i\theta}$, and one finds that the phase starts to fluctuate wildly with changes of the gauge field for $\mu\ge m_\pi/2$  at low temperature \cite{Gibbs:1986xg}.  This behavior seems precocious, since at $T=0$, nonzero $\mu$ can have no effect on the free energy until $\mu\ge m_N/3 > m_\pi/2$, where $m_N$ is the nucleon mass.  The phenomenon was clearly explained by Splittorff and Verbaarschot  \cite{Splittorff:2006fu,Splittorff:2007ck} (SV) who noted that for two degenerate flavors, $|\Delta(A_\mu,\mu)|$ correctly describes the fermion determinant with a chemical potential $\mu$ for {\it isospin} - namely $+\mu$ for the up quark and $-\mu$ for the down quark.  Such a system will exhibit Bose-Einstein condensation of pions, with  free energy becoming rapidly negative for $\mu> m_\pi/2$, as the pion is the lightest state carrying isospin.  From chiral perturbation theory they derive the estimate (in a continuum Euclidean spacetime volume $V$, for $\mu\ge m_\pi/2$)
 \beq
\frac{ \int [dA]\, \left[e^{-S_{YM}[A] }\left\vert\det\left(\slashed{\partial}-m-\mu\gamma_1\right)\right\vert^2 \,e^{2i\theta}\right]}
{ \int [dA] \, \left[e^{-S_{YM}[A] }\left\vert\det\left(\slashed{\partial}-m-\mu\gamma_1\right)\right\vert^2\right]}\simeq \frac{2\mu F_\pi/m_\pi^2}{\sqrt{2\pi V F_\pi^4}}\,e^{-2V F_\pi^2\mu^2(1-m_\pi^2/4\mu^2)^2}\ ,
\eqn{SV}\eeq
where $F_\pi$ is the pion decay constant. This result shows that the phase $\theta[A]$ is responsible for cancellations in the integral that lead to this ratio going to zero exponentially with the volume for $ \mu > m_\pi/2$.   The sign problem can be thought of as being necessary to keep pions $out$ of the ensemble for nonzero baryon number. 
 
A complementary obstacle is seen when studying the baryon spectrum  in a canonical formulation of lattice QCD, at zero chemical potential.  A typical approach to compute the mass $M_B$ of the ground state with baryon number $B$ is to use Monte Carlo methods to measure the correlation function $C_B(\tau)=\vev{\CO(0,\tau;A_\mu)}$, where $\CO=G(0,\tau;A_\mu)^{3B}$, $G$ being the  quark propagator from time $t=0$ to $t=\tau$, with color, flavor, and spin indices appropriately contracted to produce a state with the desired quantum numbers. This correlation function must behave as $C_B(\tau)\propto \exp(-M_B\tau)$ at large $\tau$ and one can therefore extract $M_B$ as the limit of $ M_B =  -\ln C_B(\tau)/\tau$ as $\tau \rightarrow \infty$. This  procedure seems straightforward and innocent of any sign problem since we have set $\mu=0$ and the fermion determinant is real and positive.  However,  one finds that a Monte Carlo measurement of $C_B(\tau)$ is very noisy at late time.  A particularly simple explanation was given by Lepage \cite{Lepage:1989hd}: he pointed out that the variance in the measurement of $C_B$ is given by the average  $\sigma^2=\vev{G(0,\tau;A_\mu)^{3B}G^\dagger (0,\tau;A_\mu)^{3B}}$, corresponding to the propagation of $3B$ quarks and $3B$ antiquarks, with a ground state consisting of $3B$ pions.  Therefore one would expect $\sigma^2\propto \exp(-3 B m_\pi \tau)$ at late time, with the signal-to-noise ratio for the Monte Carlo measurement of $C_B$ being proportional to $\exp(-3\tau(M_B/3 - B m_\pi/2))$, which vanishes exponentially fast since the mass of $B$ pions is less than $2/3$ the mass of the $B$-nucleon ground state.  Thus eventually the signal will always be overwhelmed by noise, and again it is due to the pion being lighter per constituent quark than the nucleon, just as we saw in the grand canonical example. If one works at fixed baryon density $\rho$ with  $B = V\rho$, where $V$ is the volume, one sees that the noise problem grows exponentially with the volume, as one would expect from the problem encountered in the grand canonical approach, \eq{SV}.
 
Savage has extended the analysis to look at higher moments of the distribution function for $C_B$ \footnote{M. Savage (private communication, 2010).}.  Even moments all involve equal numbers of quarks and antiquarks, and therefore fall off exponentially with a rate determined by the pion mass.  However odd moments involve expectations of operators with a net baryon number, and fall off more quickly, relative to $\sigma$ - exponentially with the nucleon mass.  With odd moments vanishing faster with $\tau$ than even moments, one can conclude that the probability distribution for $C_B$ evolves exponentially fast at late time to a  symmetric distribution with vanishing mean, a manifestation of the sign problem that in either approach is due to the existence of the light pion.
  
There have been a number of microscopic analyses of the QCD sign problem (for example, Ref.~\cite{Splittorff:2006fu}), as well as general proposals to modify the conventional approach to simulating the Feynman path integrals (such as the meron cluster approach, Ref.~\cite{Chandrasekharan:1999cm}).  Here we will instead pursue further the connection between the sign problem and chiral symmetry breaking, the origin of the pion's small mass.

 \section{The large-$N$ NJL model in three dimensions}
 We consider a modified version of the original NJL model \cite{Nambu:1961tp}, with $N$ flavors and reduced from four to three Euclidean dimensions:
 \beq
\CL = N\left(\mybar\psi_a(\slashed{\partial}-m)\psi_a- \frac{g}{2}\left[(\mybar\psi_a\psi_a)^2 + (\mybar\psi_a i\gamma_5\psi_a)^2\right]\right)\ .
\eqn{NJL}\eeq
Our convention is that $a,b, \ldots$ are flavor indices  summed over $1,\ldots,N$,  the three-dimensional (3D) coordinate indices are $i,j,\ldots$, summed over $1,2,3$, while Greek indices $\mu,\nu, \ldots$ are summed over four-dimensional (4D) coordinates $1,\ldots,4$. Thus $\slashed{\partial} = \gamma_i\partial_i$, for example. The gamma matrices are the usual $4\times 4$ matrices used in 4D, not the $2\times 2$ matrices appropriate for 3D, and so the Lagrangian represents $2N$ flavors of 3D Dirac fermions.  However, we will generally use 4D nomenclature - in particular, in the limit $m\to 0$ this theory has a $U(1)$ chiral symmetry in 4D, which becomes a flavor symmetry in 3D; despite the fact that there is no notion of chiral symmetry in 3D, we will continue to refer to this global $U(1)$ symmetry as a chiral symmetry.

In 3D this model still exhibits the key feature of the original NJL model,  the phenomenon of spontaneous chiral symmetry breaking that gives rise to a Goldstone boson we will call the pion.  We will analyze this theory in a $1/N$ expansion using techniques similar to those used in the Gross-Neveu model, which is formulated in two-dimensional (2D) \footnote{Although the theory given in Eq. (3) is often referred to as the Gross-Neveu model, it is not asymptotically free and chiral symmetry breaking requires a critical value for the coupling $g$ as in the original NJL model in 4D, and unlike the original Gross-Neveu model at large-$N$ in 2D; we will continue referring to it as the 3D NJL model at large $N$.}. This theory has been reviewed  in \cite{Hands:1992be,Rosenstein:1990nm} and has been studied numerically in \cite{Hands:2001cs}.

\subsection{The $\sigma/\pi$ formulation}

The conventional treatment of the theory \eq{NJL} is to introduce auxiliary fields $\sigma$ and $\pi$ to obtain a bilinear fermion action,
\beq
\CL =N\left( \frac{1}{2g}\left(\sigma^2+\pi^2\right) +\mybar\psi_a \left[\slashed{\partial}-m+\sigma + i\pi\gamma_5\right]\psi_a\right)\ .
\eqn{SP}\eeq
In this formulation, the $\sigma$ and $\pi$ fields are singlets under the $SU(N)$ flavor symmetry, while $\phi = (\sigma+i\pi)/\sqrt{2}$ transforms linearly under the chiral $U(1)$ symmetry.  With the above normalization, $N$ counting is simple: every vertex and every fermion loop brings a factor of $N$, every propagator a factor of $1/N$.  Loops that include scalar propagators do not give a factor of $N$, since the mesons do not carry the $N$ flavors.

\subsubsection{No sign problem}

An interesting feature of this theory is that at finite chemical potential the fermion determinant is real, and for even $N$, it is positive. To see this, note that we can define a  real  symmetric charge conjugation matrix $C$ that satisfies $C^2=1$, $C\gamma_i C=\gamma_i^*$ for $i = 1,2,3,$ and $C\gamma_5 C = -\gamma_5^*$.  For example, we can take $\gamma_i = \sigma_1\times \sigma_i$, $\gamma_5 = \sigma_3\times 1$ with $C=\sigma_2\times\sigma_2$.  Then the fermion operator for a single flavor in the grand canonical formulation satisfies $D^* = CDC$, where $D= (\slashed \partial -m + \sigma + i\pi\gamma_5 + \mu \gamma_1)$.  It follows that complex eigenvalues of $D$ come in conjugate pairs, while real eigenvalues can be unpaired.  Thus $\det D$  for the $N$-flavor theory, \eq{SP}, equals a real number raised to the power $N$ and is positive for even $N$. This implies that there is no sign problem obstacle to simulating this theory at finite density using Monte Carlo methods \cite{Hands:1995jq}.

\subsubsection{Chiral symmetry breaking}

Integrating out the fermions gives rise to the effective action  $S_\text{eff} = N S(\sigma,\pi)$, where
\beq
S[\sigma,\pi] =\int {\rm d}^3x  \left[ \frac{ 1}{2g } (\sigma(x)^2+\pi(x)^2) -\Tr\ln(\slashed{\partial}-m+\sigma(x)+i\pi(x)\gamma_5)\right] \ .
\eqn{ssp}\eeq
The large-$N$ expansion is equivalent to the semiclassical expansion, and the vacuum is characterized by the classical solution that minimizes the action $S$.  We can readily compute $S(\sigma,0)$ with $\sigma=$ constant using dimensional regularization and minimal subtraction (MS) \footnote{In this theory minimal subtraction MS means no subtraction, as there are no logarithmic divergences and hence no $1/(d-3)$ poles}.  Up to an irrelevant additive constant we find
\beq
S(\sigma,0) = V  T\left[\frac{ \sigma^2}{2g}+\frac{\left[(\sigma-m)^2\right]^{3/2}}{3\pi}\right]\ ,
\eqn{Sval}\eeq
where we have put the system in a box of spacetime volume $V  T$.  Defining the chiral symmetry breaking minimum to be at $\vev{\sigma} = f$, when $m=0$, we obtain
\beq
\frac{\partial S(\sigma,0)}{\partial\sigma}\Biggl\vert_{\substack{m=0\\ \sigma=f}}= 0\quad \Longrightarrow\quad g=-\frac{\pi }{f}\ ,
\eqn{Cval}\eeq
with $f>0$. With this value for $g$ (which is renormalization scheme dependent), we find for nonzero quark mass $m>0$
\beq
S(\sigma,0) = V  T\left[-\frac{ \sigma^2f}{2\pi}+\frac{\left[(\sigma-m)^2\right]^{3/2}}{3\pi}+\CS_0\right]\ ,
\eqn{Sval}
\eeq
with minimum at
\beq
\vev{\sigma} = \frac{f}{2}\left[1+\sqrt{1+4\eta} + 2 \eta\right] \ ,\qquad\eta\equiv\frac{m}{f}\ ,
\eqn{sigmavev}\eeq
which shows how the chiral symmetry breaking minimum depends on the explicit quark mass. We also choose the constant, 
\beq
%\CS_0 = -\frac{f^3}{24 \pi}\left[2\sqrt{2}\left(1+ 2\eta+\sqrt{1+4\eta}\right)^{3/2}-3\left(1+ 2\eta + \sqrt{1+4\eta}\right)^2 \right] \,
\CS_0 &=& \frac{f^3}{12\pi}\left[(1+4\eta)^{3/2} +(1+6\eta+6\eta^2)\right]\ ,
\eeq
such that the action vanishes in the chiral symmetry breaking vacuum.

Next we expand the effective action $S$ to second order in spacetime dependent fluctuations of the $\sigma$ and $\pi$ fields,
\beq
S(\sigma,\pi) =
S(\vev{\sigma},0) +    \frac{1}{2}  V^2  T^2 \int \frac{d^3k}{(2\pi)^3}\left[D_\sigma(k)\delta\sigma(k)\delta\sigma(-k) + D_\pi(k) \delta\pi(k)\delta\pi(-k) \right] +\ldots
 \eqn{Sexp}\eeq
where $D_\sigma$ and $D_\pi$ are readily computed in the MS scheme from one-loop diagrams in the background $\vev{\sigma}$.  The constituent quark mass M is given by
\beq
M \equiv  \vev{\sigma}-m = \frac{f}{2}\left[1+\sqrt{1+4\eta} \right]\ .%\qquad \eta\equiv \frac{m}{f}\ ,
\eqn{Mval}\eeq
Since there is no confinement in this theory, $M$ is the mass of the lightest fermionic excitation and $m$ is the current quark mass. With the fermion propagator, $G(p) \equiv(-i\slashed{p}+M)^{-1}$, we find,
\beq
D_\sigma(k) &=&\frac{1}{g} + \int\frac{d^3q}{(2\pi)^3} \Tr\left[ G(q+k/2) G(q-k/2)\right]=-\frac{f}{\pi }+\frac{\left(4 M^2+k^2\right) \cot ^{-1}\left(2 M/k\right)+2 M k}{2 \pi  k}\ ,\cr&&\cr
D_\pi(k) &=&\frac{1}{g} - \int\frac{d^3q}{(2\pi)^3} \Tr\left[ \gamma_5 G(q+k/2) \gamma_5 G(q-k/2)\right]= \frac{2M-2f+k\cot^{-1}\left(2M/k\right)}{2\pi}\ .
\eqn{spprops}\eeq
To leading order in $1/N$, these functions determine the $\sigma$ and $\pi$ dispersion relations, and their masses are defined by the location of zeros in $D_\sigma$ and $D_\pi$, respectively. One finds $m_\sigma^2 = (2 f)^2$ in the chiral limit. However, the pole sits at the beginning of the two-fermion branch cut, for $m>0$ we find $m_\sigma>2 M$, and so the $\sigma$ field is unstable.   Only at subleading order in $1/N$ would one see the branch cut appear in $D_\sigma$ for $\sigma\to \pi\pi$ decay, as that entails an additional quark loop.  A chiral expansion for the pion mass yields
\beq
m_\pi^2 = 4 f m\left(1-\frac{1}{3} \eta + \frac{31}{45}\eta^2 -\frac{1654}{945}\eta^3 + \ldots\right)\ ,
\eeq
and near the pion pole, $k^2=-m_\pi^2$, one finds for the pion propagator
\beq
\frac{1}{N D_\pi(k)} \simeq \frac{1}{N}\frac{Z_\pi}{k^2+m_\pi^2}\ , \qquad Z_\pi = 4\pi f\left(1+\frac{1}{3}\eta - \frac{4}{15}\eta^2 + \ldots\right)\ .
\eqn{zprop}
\eeq
The positivity of $m_\sigma^2$ and $m_\pi^2$ for $m>0$ shows that we found the correct (e.g. stable) vacuum.

\subsection{The $A/V$ formulation}
An alternative formulation of the theory follows if one first uses the  Fierz identity 
\beq
\left[\delta_{ij}\delta_{kl} - (\gamma_5)_{ij}(\gamma_5)_{kl}\right] = \half\left[ (\gamma_\mu)_{il}(\gamma_\mu)_{kj} - (\gamma_\mu\gamma_5)_{il}(\gamma_\mu\gamma_5)_{kj}\right]
\eqn{Fierz}\eeq
to rearrange the four-fermion interaction in \eq{NJL} before introducing auxiliary fields. The Lagrangian becomes 
\beq
\CL = N\left(\mybar\psi_a(\slashed{\partial}-m)\psi_a+ \frac{g}{4}\left[(\mybar\psi_a\gamma_\mu \psi_b)(\mybar\psi_b\gamma_\mu\psi_a) - (\mybar\psi_a\gamma_\mu\gamma_5\psi_b)(\mybar\psi_b\gamma_\mu\gamma_5\psi_a )\right]\right)
\eqn{NJLfierzed}\eeq
where $\slashed{\partial} = \gamma_i\partial_i$ with $i$ summed over $i=1,2,3$, while the $\gamma_\mu$ matrices are summed over $\mu=1,\ldots,4$.   This formulation invites the introduction of $N\times N$ matrix valued vector and axial vector auxiliary fields $V$ and $A$, giving the equivalent theory
\beq
\CL =N\left( \frac{1}{g}\, \Tr\left(V_\mu V_\mu+A_\mu A_\mu \right) +\mybar\psi \left[\slashed{\partial}-m+i\slashed{V} + \slashed{A}\gamma_5\right]\psi\right)\ .
\eqn{VA}\eeq
$N$ counting in this theory is different from the $\sigma/\pi$ formulation, since the $A$ and $V$ meson fields are $N\times N$ matrices.  In fact, the $N$ counting here is identical to that of large-$N$ QCD, and it is convenient to employ 't Hooft's double-line notation for the mesons. As in QCD, the order of a graph without external legs is given by 
$N^\chi$, where $\chi$ is the Euler characteristic of the surface defined by the graph, and so to leading order one only need consider planar graphs with a minimal number of closed fermion loops.  However this class of graphs is much simpler than in QCD, since the $A$ and $V$ mesons have no cubic or quartic interactions, unlike gluons.

\subsubsection{A QCD-like  sign problem}

In this $A/V$ formulation, the fermion matrix at finite chemical potential is given by $D(\mu) = \left[\slashed{\partial}-m+i\slashed{V} + \slashed{A}\gamma_5+\mu\gamma_1\right]$, which is similar in structure to the QCD Dirac matrix with nonzero chemical potential $\mu$ (with the $N$ flavors playing the role of color) and its determinant is similarly complex.  In fact, as in QCD we can make an SV argument  \cite{Splittorff:2006fu,Splittorff:2007ck} about the phase of the fermion determinant by considering two degenerate families (e.g. $2N$ fermions) so that the chiral symmetry is enlarged from $U_L(1)\times U_R(1)$ to $U_L(2)\times U_R(2)$.  The fermion determinant in the case of a quark number chemical potential is $(\det D(\mu))^2$ while with an isospin chemical potential it is  $\vert \det D(\mu)\vert^2$, the difference between the two being the phase $e^{2i\theta}$.  In the latter case there is a transition to a pion-condensed state at $\mu=m_\pi/2$ just as in QCD, and so the SV argument leads to a similar formula as in \eq{SV}.

 It is remarkable that a theory with a sign problem so similar to that of QCD is known to have a formulation that has no sign problem, the $\sigma/\pi$ formulation of the previous section.

\subsubsection{Chiral symmetry breaking}
\label{ChiSB}

One cannot see chiral symmetry breaking in the $A/V$ formulation in a mean field formalism, as there is no fundamental field with the right quantum numbers to play the role of the $\mybar\psi\psi$ condensate - again, as in QCD.   Instead one has to consider the Schwinger-Dyson equation for the fermion propagator, which is exact in leading order in $1/N$ since vertex and meson propagator corrections occur at higher order.  We take the full canonically normalized fermion propagator to equal
\beq
G(p) = \frac{1}{-i\slashed{p} Z(p)+ M(p)} \ ,
\eeq
which satisfies the integral equation
\beq
-i\slashed{p}Z(p) +M(p)&=& -i\slashed{p} -m -\frac{g}{2} \int\frac{{\rm d}^3k}{(2\pi)^3 }\left[-\gamma_\mu\left(\frac{1 }{-i\slashed{k}Z(k)+ M(k)}\right)\gamma_\mu + \gamma_\mu\gamma_5\left(\frac{1 }{-i\slashed{k}Z(k)+ M(k)}\right)\gamma_\mu \gamma_5\right]\cr&&\cr
 &=&
 -i\slashed{p} -m +4g \int\frac{{\rm d}^3k}{(2\pi)^3 }\frac{M(k)}{k^2 Z(k)^2+ M(k)^2} \ .
  \eeq
From this one finds (using the MS renormalization scheme as before) that $Z(k)=1$ and $M(k)=M$ is a constant satisfying
\beq
M =  -m+4g \int\frac{{\rm d}^3k}{(2\pi)^3 }\frac{M}{k^2  + M^2} = -m - \frac{g M^2}{\pi}\ .
\eeq
which has the solution
\beq
M =  \frac{\pi}{2g} \left(-1 \pm \sqrt{1-\frac{4gm}{\pi}}\right)    \ .
\eeq
Using the renormalization condition that the dynamical fermion mass equals $f$ in the the chiral limit, $m=0$, gives $g=-\pi/f$, as in \eq{Cval} and the solution for $M$
\beq
M = \frac{f}{2} \left(1 + \sqrt{1+4\eta}\right) \qquad \eta \equiv \frac{m}{f}   \ ,
\eqn{self}\eeq
which is the same as the value derived  in the $\sigma/\pi$ formulation for the dynamical fermion mass,  \eq{Mval}.

\begin{figure}[t]
\centerline{\includegraphics[width=3.0in]{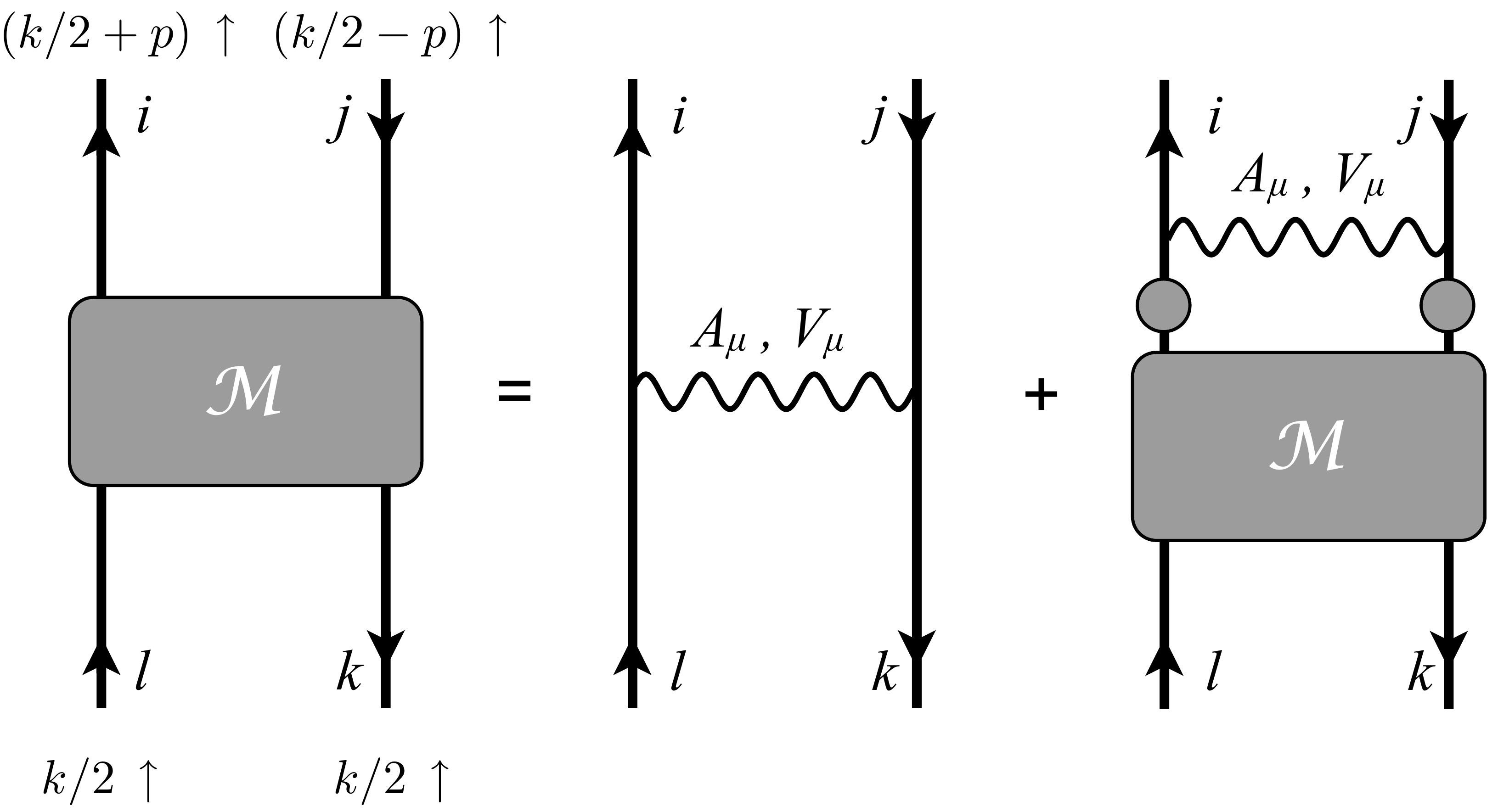}}
\caption{\it A graphical representation of the integral equation \eq{inteq} for the four-point correlator in the $A/V$ formulation for an incoming fermion/antifermion pair of one flavor and an outgoing pair of another.  Dirac indices are labeled.}
\label{fig:fourpoint}
\end{figure}

\begin{figure}[t]
\centerline{\includegraphics[width=3.5in]{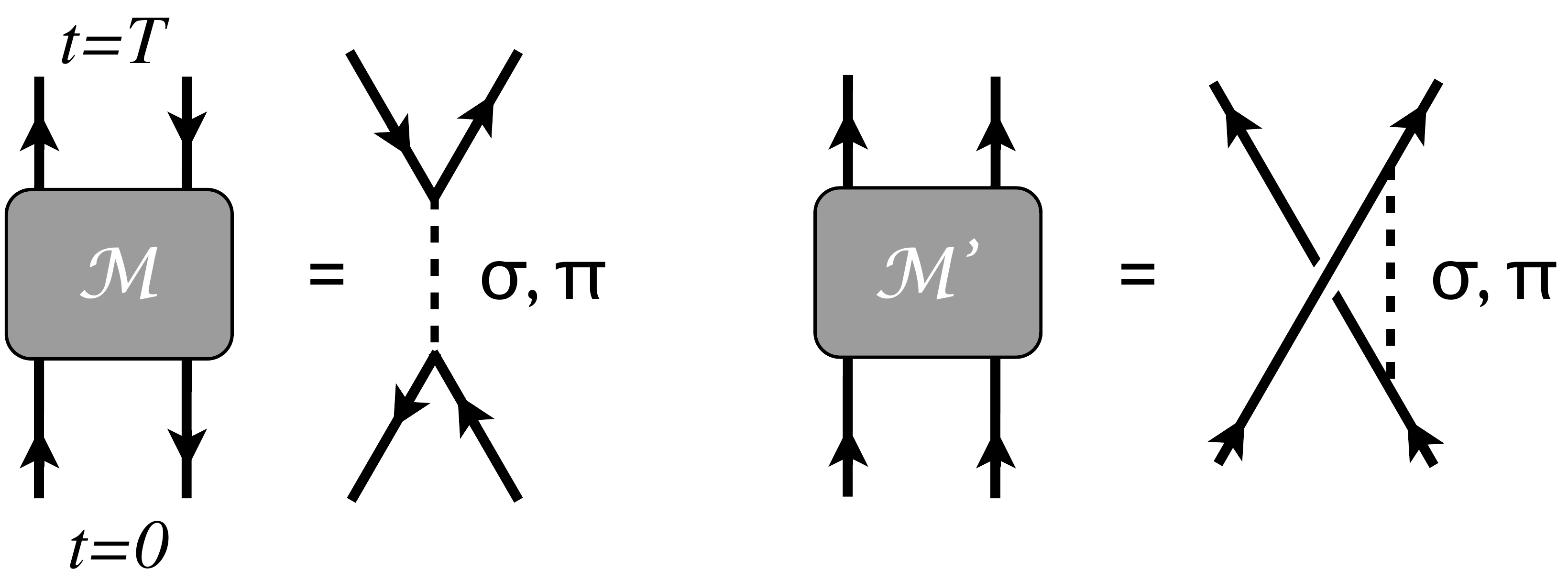}}
\caption{\it In the $A/V$ formulation at leading order in $1/N$, interaction between a valence fermion/antifermion pair ($\CM$) or a valence fermion pair ($\CM'$)  is equivalent to $s$- or $u$-channel exchange respectively of single $\sigma$ and $\pi$ mesons in the $\sigma/\pi$ formulation, where the fermions have mass $M$ arising from nonzero $\vev{\sigma}$. }
\label{fig:AVequiv}
\end{figure}

In the $A/V$ formulation the $\sigma$ and $\pi$ mesons appear as  a fermion and antifermion pair bound together by strong $A_\mu$ and $V_\mu$ vector meson exchange, much the same way as mesons arise in large-$N$ QCD.  To see them we compute the connected four-point function $\CM$ for a fermion/antifermion pair of flavors $\ell,k$, each with 3-momentum $k/2$, to scatter into a fermion/antifermion pair of flavors $i,j$ with 3-momenta $k/2\pm p$.  To leading order in $1/N$, $\CM$ obeys the integral equation, shown graphically in Fig.~\ref{fig:fourpoint},
\beq
\CM_{ij;kl}(k,p) &=&  \frac{ g N}{2}\left[- (\gamma_{\mu})_{il}(\gamma_{\mu})_{kj}+(\gamma_{\mu}\gamma_{5})_{il} (\gamma_{\mu}\gamma_{5})_{kj} \right] \cr&&+ \frac{g N^2}{2}
\int \frac{d^3q}{(2\pi)^3} G(q+k/2)_{ab}\CM_{bc;kl}(k,q) G(q-k/2)_{cd} [- (\gamma_{\mu})_{ia}(\gamma_{\mu})_{dj}+(\gamma_{\mu}\gamma_{5})_{ia}(\gamma_{\mu}\gamma_{5})_{dj} ] \cr
&=&- {g N} \left[\delta_{ij}\delta_{kl}-(\gamma_5)_{ij} (\gamma_5)_{kl}\right] \cr&& -g N^2
\int \frac{d^3q}{(2\pi)^3} G(q+k/2)_{ab}\CM_{bc;kl}(k,q)G(q-k/2)_{cd} \left[\delta_{ij}\delta_{da}-(\gamma_5)_{ij} (\gamma_5)_{da}\right] \ .\eqn{inteq}\eeq
where $G(p)$ is the  free fermion propagator with  dynamical mass $M$, and we used a Fierz identity to replace the vector and axial vector gamma matrices by scalar and pseudoscalar.  The solution to this equation is 
\beq
\CM_{ij;kl}(k)  =-{N}\left( \frac{\delta_{ij}\delta_{kl}}{D_\sigma(k)} +  \frac{(i\gamma_5)_{ij}(i\gamma_5)_{kl}}{D_\pi(k)} \right) = -N^2\left(  \delta_{ij}\delta_{kl} G_\sigma(k) +  (i\gamma_5)_{ij}(i\gamma_5)_{kl}G_\pi(k) \right) \ ,
\eqn{cmval}\eeq
where we dropped the label $p$ (as our solution is independent of $p$), $D_{\sigma,\pi}$ are given in \eq{spprops}, and $G_{\sigma,\pi}=N/D_{\sigma,\pi}$ are the full meson propagators.  Thus we see that (up to a sign) the interaction between  valence fermion and   antifermion  via $t$-channel exchange of $A_\mu$ and $V_\mu$ mesons is exactly equivalent to an annihilation diagram in the $\sigma/\pi$ formulation, with a single meson in the $s$-channel;  similarly, the interaction between two valence fermions or two valence antifermions in the $A/V$ theory is equivalent to a single meson exchange in the $u$ channel in the $\sigma/\pi$ theory (Fig.~\ref{fig:AVequiv}).  This equivalence will allow us to use the simpler $\sigma/\pi$ theory to calculate the cumulants for the $A/V$ theory.

%%%%%%%%%%%
%%%%%%%%%%%

%%%%%%%%%%%
%%%%%%%%%%%

\section{The probability distribution of the fermion propagator}
If $X[\phi]$ is a functional of a stochastic field $\phi$ corresponding to an observable  --- such as a correlation function --- we define the normalized probability density function (pdf)  for $X$ to be the path integral
\beq
\CP(x) =\CN \int [{\rm d}\phi]\, e^{-S[\phi]} \,\delta(X[\phi]-x)\ .
\eqn{pdf}\eeq
where we assume $S[\phi]$ is real. If one were to sample an ensemble of $\phi$ configurations according to the distribution $e^{-S[\phi]}$, the values of $X[\phi]$ would be distributed according to $\CP$, making its relevance to Monte Carlo simulations evident.  For an accurate estimate of $\vev{X}$ from a reasonable number of samples, one would like $\CP(x)$ to be sharply peaked around its mean.  However, one might  find a very broad distribution centered about a mean close to zero, making an accurate estimate of the mean, without a huge number of samples, very difficult, such as is the case with baryon propagators in QCD. Alternatively, one might find a heavy-tailed distribution for which very rare events make a significant contribution to the mean, resulting in very noisy and often misleading estimates of $\vev{X}$ from a finite sample.   This latter situation is indicative of what is called ``an overlap problem," which occurs when $e^{-S[\phi]}$ is peaked far from the field configurations that provide support for nonzero $X(\phi)$.   With some knowledge about the nature of the tail of the distribution, it may be possible to use statistical methods to greatly improve the determination of $\vev{X}$ \cite{Endres:2011jm}.

The pdf given in \eq{pdf} is a difficult quantity to analyze using field theoretic methods because of the singular nature of the delta function; instead, we consider the characteristic function (cf) $\Phi_X(s)$, which is just the Fourier transform of the pdf:
\beq
\Phi_X(s) = e^{-W(s)} =  \CN\int [{\rm d}\phi]\, e^{-S[\phi] + i s X[\phi]} \ .
\eqn{rephi}\eeq
This is a useful formulation because $W(s) =- \ln \Phi_X(s)$ is on the one hand given by the connected Feynman diagrams \footnote{Since the functional $X[\phi]$ is typically nonlocal, by connected Feynman diagrams we mean neither conventional Feynman diagrams, nor conventionally connected.  For example,  expanding $X[\phi]$ to second order in $\phi$ might take the form $\left(\int d^3x F(x) \phi(x)\right)^2$ for some function $F$; this is treated as a fundamental 2-point vertex in our discussion.} of the modified action $S[\phi]-isX[\phi]$, while, up to an irrelevant additive constant, it is also the generating function for the cumulants of $X$:
\beq
W(s) = -\sum_{n=1}^\infty \frac{(i s)^n}{n!}\, \kappa_n\ .
\eqn{cumulantsexp}\eeq
Here, $\kappa_n$ is the $n$th cumulant, with $\kappa_1=\vev{X}\equiv\mu$ being the mean of $\CP(x)$,  $\kappa_2=(\vev{X^2}-\vev{X}^2)\equiv\sigma^2$ being the variance, etc.  

This procedure has to be modified slightly when dealing with a complex observable, where we replace \eq{rephi} by
\beq
\Phi_X(s, \bar s) = e^{-W(s, \bar s)} =  \CN\int [{\rm d}\phi]\, e^{-S[\phi] + i (s X[\phi]+\bar s \bar X[\phi])} \ .
\eqn{compphi}\eeq
where $s$ is now complex and the bar indicates complex conjugation.  Now, $\ln W(s,\bar s)$ has a double expansion in both $s$ and $\bar s$,
\beq
W(s, \bar s) = -\sum_{m,n=1}^\infty \frac{(i s)^m(i \bar s)^n}{m!\,n!}\, \kappa_{m,n}\ .
\eqn{compkappa}\eeq
with $\kappa_{n,m} = \bar\kappa_{m,n}$.  For example, 
\beq
\kappa_{1,0} = \vev{X}\ ,\qquad 
\kappa_{1,1} = \vev{|X|^2} - |\vev{X}|^2\ ,\qquad
\kappa_{2,0} = \vev{X^2} - \vev{X}^2\ ,\qquad \ldots
\eqn{egcompkappa}\eeq

In what follows we show how to compute the cumulants of the correlation function of a single fermion with zero 2-momentum for a time extent $\tau$.  This is the sort of measurement one would perform in a Monte Carlo simulation in order to determine the mass of the fermion.  In particular we take for $X[\phi]$
\begin{subequations}
\begin{align}
 Y_\Gamma&=\ln\left[\frac{1}{V}\expect{\bfp=0,t=\tau/2\,}{\Tr \Gamma\left(\frac{1}{\slashed{\partial} -m+\sigma + i \pi\gamma_5}\right)}{\bfp=0,t=-\tau/2}\right]\quad &\sigma/\pi\text{ formulation,}\eqn{corrsp}\\
 X_\Gamma&=\frac{1}{V}\expect{\bfp=0,t=\tau/2\,}{\Tr \Gamma\left(\frac{1}{\slashed{\partial} -m+i\slashed{V} + \slashed{A}\gamma_5}\right)}{\bfp=0,t=- \tau/2}\quad &A/V\text{ formulation,}\eqn{corrAV}
 \end{align}
 \end{subequations}
 where $\Gamma$ is some Dirac matrix of our choosing. Our choice to look at the log of the propagator in the $\sigma/\pi$ formulation makes the later analysis simpler. Note that in the definitions of our observable $X_\Gamma$ and $Y_\Gamma$ we use canonically normalized fermion propagators, without the factor of $1/N$. 
 Measuring the expectation value of this correlator is a procedure for determining the mass $m_f$ of the lightest fermion state through the formula
 \beq
\lim_{\tau\to\infty} \ -\frac{1}{\tau} \ln \vev{X_\Gamma} = \lim_{\tau\to\infty} \ -\frac{1}{\tau} \ln \vev{e^{Y_\Gamma}} =m_f\ ,
\eqn{plat}\eeq
provided that $\Gamma$ does not project out this state.  Of course, we already know analytically that $m_f=M$, with $M$ given in \eq{Mval}, but by calculating the cumulants for this observable we will establish how difficult it would be  to determine the fermion mass by  numerical Monte Carlo methods using the two formulations of the NJL model, and why.  In particular we will show that in the QCD-like $A/V$ formulation, the pdf for $X_\Gamma$, at late time,  looks as one would expect from the Lepage-Savage picture: a broad distribution that is nearly symmetric about zero with an exponentially small mean.  In contrast, the pdf for the physically equivalent $\sigma/\pi$ formulation looks heavy-tailed and close to log-normal at late time.  Thus, a Monte Carlo study of this theory, without a sign problem, still faces an overlap problem and significant numerical challenges, but is perhaps amenable to a cumulant expansion as introduced for a similar system in Refs.~\cite{Lee:2011sm,Endres:2011jm,Endres:2011mm}. This theory has also been successfully investigated recently using the ``fermion bag approach" \cite{Chandrasekharan:2009wc,Chandrasekharan:2011mn,Chandrasekharan:2012va}.

\subsection{Noise distribution in the $A/V$ formulation and the Lepage-Savage analysis}

\subsubsection{The $\kappa_{1,0}$ and $\kappa_{1,1}$ cumulants}

We begin by computing the cumulants for measurements of the fermion propagator $X_\Gamma$  in the $A/V$ formulation; since this observable is complex, we use the formalism in \eqs{compphi}{egcompkappa}.  From the above discussion, the $\kappa_{mn}=\bar\kappa_{nm}$ cumulant for $X_\Gamma$ is given by the sum of connected graphs with $m$ copies of $X_\Gamma$ and $n$ copies of its complex conjugate, $\bar X_\Gamma$.  We will refer to  these as  valence fermion and valence antifermion propagators, respectively; at leading order in $1/N$ there are no sea quark loop contributions.  With our definition of $X$, there are no factors of $1/N$ from valence fermion propagators, nor factors of $N$ from meson coupling to valence fermions, nor is annihilation of a valence fermion with a valence antifermion allowed.  The computation of $\kappa_{m,n}$ involves graphs with net fermion number $(m-n)$, and according to the Lepage-Savage analysis, we expect 
\beq
\kappa_{m,n}\propto e^{-\left[(m-n)M + n m_\pi\right]\tau}\qquad (m\ge n)\ .
\eqn{LS}\eeq
We will see that this expectation is born out in the $A/V$ formulation, which is the one with a QCD-like sign problem at nonzero chemical potential.

We should not compute graphs contributing to the valence fermion self-energy, but rather use the nonperturbative solution to the Schwinger-Dyson equation, replacing the mass term $(-m)$ in the fermion propagator by $M$ from \eq{self}.  It is convenient to have this propagator in a mixed $\{t,\bfp\}$ representation:
\beq
\expect{\bfp',t'}{\frac{1}{\slashed{\partial} + M}}{\bfp,t} \equiv (2\pi)^2\delta^2(\bfp'-\bfp)\widetilde G(\bfp,t'-t)\ ,
\eeq
with
\beq
\widetilde G(\bfp,t)=\int\frac{{\rm d}\omega}{2\pi}\frac{e^{-i\omega t}}{-i \omega\gamma_1 - i \bfp\cdot{\mathbf{\gamma}}+M}=e^{-\omega_p|t|} \frac{\omega_p\epsilon(t) \gamma_1 + i\bfp\cdot\mathbf{\gamma}+ M}{2\omega_p}\ ,
 \eeq
where $\omega_p = \sqrt{|\bfp|^2 + M^2}$.
Note that
\beq
\widetilde G (\mathbf{0},t) = e^{-M|t|}\left( \frac{1+\epsilon(t)\gamma_1}{2} \right)\ .
\eqn{gtilde0}
\eeq 
is proportional to a projection operator.  Since the dynamical mass $M$ includes all nonperturbative contributions to the fermion self-energy, it follows that the first cumulant for $X_\Gamma$ is just
\beq
\kappa_{1,0} = \Tr\left[\Gamma \widetilde G (\mathbf{0},\tau)\right] = z e^{-M\tau}=\vev{X_\Gamma}\ ,
\eeq
where $z \equiv \Tr\left[\Gamma \left(\frac{1+\gamma_1}{2}\right)\right]$ is the wave-function overlap of our chosen observable with the physical fermion state. 
Thus a Monte Carlo simulation that correctly estimates the value of $\vev{X_\Gamma}$ will correctly determine the fermion mass to equal $M$ by means of the formula  \eq{plat}, provided that $\Gamma$ is chosen so that $z$ is nonvanishing.

As we are interested in how difficult a Monte Carlo determination of $\vev{X_\Gamma}$ might be,  we turn next to the variance  $\kappa_{1,1}$. At leading order in $1/N$, the sum of diagrams for $\kappa_{1,1}$ is given by attaching $\widetilde G$ propagators at zero spatial momentum to the legs in the first diagram in Fig.~\ref{fig:fourpoint}, with ends $k,l$ at time $t=-\tau/2$ and ends $i,j$ at time $\tau/2$, and contracting the Dirac indices of each valence fermion line with $\Gamma$.  Using our result for the four-point function $\CM$ in \eq{cmval} (but dividing by $N^2$, since our  $X_\Gamma$ is the propagator for a canonically normalized fermion), we find that the $D_\sigma$ part of $\CM$ is killed by the Dirac trace and only the $D_\pi$ part contributes, yielding 
\beq
\kappa_{1,1} &=& \frac{1}{V N}\int_{-\infty}^\infty {\rm d}t_1\, \int_{-\infty}^\infty {\rm d}t_2\,\Tr\Gamma \widetilde G (\mathbf{0},\tau/2-t_2)\gamma_5\widetilde G (\mathbf{0},t_2-\tau/2)\Gamma \widetilde G (\mathbf{0},-\tau/2-t_1)\gamma_5\widetilde G (\mathbf{0},t_1+\tau/2) \int\frac{{\rm d}\omega}{2\pi}\,
\frac{e^{-i\omega(t_2-t_1)}}{D_\pi(\omega)}
\cr&&\cr
&=& \int_{-\infty}^\infty {\rm d}t_1\, \int_{-\infty}^\infty {\rm d}t_2\,e^{- M(|\tau- 2t_2|+|\tau+2t_1|)} \Tr\Gamma \left(\frac{1+\epsilon(\frac{\tau}{2}-t_2)\gamma_1}{2}\right)\gamma_5\Gamma \gamma_5 \left(\frac{1+\epsilon(t_1+ \frac{\tau}{2})\gamma_1}{2}\right) \bar G_\pi(t_2-t_1)\ ,
\eeq
where
\beq
 \bar G_\pi(t) \equiv \int\frac{d^2x}{V} \, G_\pi(\bfx,t) = \frac{1}{V N}\int\frac{{\rm d}\omega}{2\pi}\,
\frac{e^{-i\omega(t)}}{D_\pi(\omega)} \simeq\frac{ Z_\pi}{2m_\pi V N} e^{-m_\pi |t|}
\eeq
with $D_\pi$, $Z_\pi$  given in \eq{spprops} and \eq{zprop}, respectively, and we approximated $D_\pi^{-1}(\omega)$ by the pion pole contribution, \eq{zprop}, ignoring the branch cut at $k^2 <  -4 M^2$. 
If we choose $\Gamma=1$, we find
\beq
\kappa_{1,1} &\simeq& \frac{ Z_\pi}{2m_\pi V N} \int_{-\infty}^\infty {\rm d}t_1\, \int_{-\infty}^\infty {\rm d}t_2\,e^{-2 M(|\frac{\tau}{2}-t_2|+|t_1+ \frac{\tau}{2}|)}\left(1+ \epsilon\left(\frac{\tau}{2}-t_2\right)\epsilon\left(t_1+\frac{\tau}{2}\right)\right)e^{-m_\pi |t_2-t_1|}\, \\
& \simeq& 
\frac{2\pi }{ m_\pi f V N } e^{-m_\pi \tau} \ , \eeq
where we have (i) assumed we are near the chiral limit with $m_\pi \ll f, M$ and (ii) assumed we are interested in late time behavior, $\tau \gg 1/M$.
Note that near the chiral limit we are finding $\kappa_{1,0}/\sqrt{\kappa_{1,1}} \propto \exp[(-M+m_\pi/2)\tau]\ll1$ at late time, indicating a severe signal-to-noise problem, in agreement with the Lepage analysis. 

 It is interesting to note that if instead we take $\Gamma=(1\pm\gamma_1)/2$ then $\kappa_{1,1}$ vanishes at this order in $1/N$. This choice of $\Gamma$ kills the pion contribution to the variance, and so we would expect a noise-free measurement in this case.   We do not expect this to persist at subleading order in $1/N$, nor do we expect that in real QCD one can decouple baryon observables  from pions so easily, but it seems worth exploring whether correlating initial and final Dirac indices of baryon operators  (as with this trace with $(1\pm\gamma_1)/2$ on valence quark lines) might be able to improve the signal-to-noise problem in real QCD computations.

\subsubsection{Power counting for higher $\kappa_{m,n}$ cumulants}

\begin{figure}[t]
\includegraphics[width=4in]{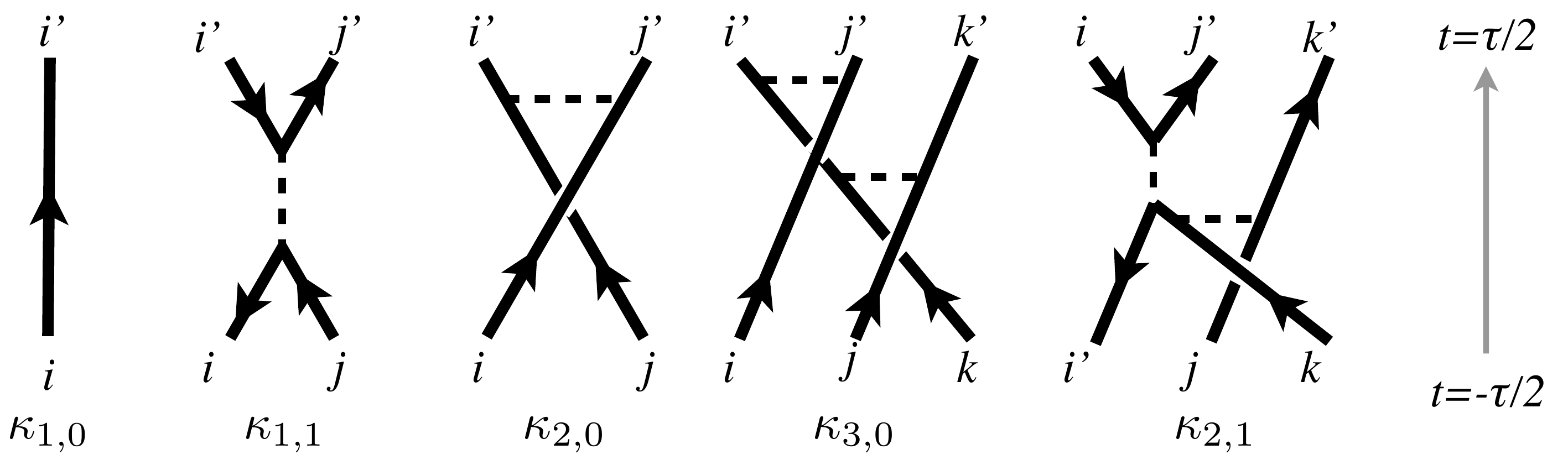}
\caption{\label{fig:AVkappas} {\it  Contributions to several low cumulants for the $A/V$ theory to leading order in $1/N$, using the $\sigma/\pi$ method to compute them as developed in Sec. \ref{ChiSB}.  Solid lines represent fermion propagators, and dashed lines are mesons. The end points are to be contracted with $\Gamma_{ii'}$, $\Gamma_{jj'}$, $\Gamma_{kk'}$.  The lightest intermediate state that can appear in a graph for $\kappa_{m,n}$ with $m\ge n$ consists of $(m-n)$ fermions and $n$ pions, and the sum of their masses determines the $\tau$ dependence of the cumulant.}}
\end{figure}

Higher cumulants can be computed for the $A/V$ formulation using the equivalent $\sigma/\pi$ diagrams as discovered in Sec. \ref{ChiSB} and shown in Fig.~\ref{fig:AVequiv}.  This is {\it not} to say that the same diagrams give the cumulants for the $A/V$ and the $\sigma/\pi$ theories.  In Fig.~\ref{fig:AVkappas} we show the diagrams for several of the lower cumulants, where solid lines are the fermion propagators $1/(-i \slashed{p} +M)$, and dashed lines are the meson propagators $G_\pi$ and $G_\sigma$ (both mesons contributing in general) with couplings $1$ or $i\gamma_5$, respectively, at the vertices.  Again, we take all incoming and outgoing 2-momenta to be zero, and there is one $\Gamma_{i i'}$ contracted with each pair of like indices in the graph.
As we have chosen a canonically normalized fermion propagator with one particular  flavor as our observable, there are no factors of $N$ at the meson vertices, nor are there any loops giving rise to factors of $N$.  However, each meson propagator costs a factor of $1/N$, and so $\kappa_{m,n}\propto (1/N)^{m+n-1}$, since we need a minimum of $(m+n-1)$ mesons to make a connected graph.  Furthermore, one can see by cutting the graphs at a fixed time that the minimum mass state that can possibly propagate in a graph for $\kappa_{m,n}$ with $m\ge n$ consists of $(m-n)$ fermions with mass $M$  and $n$ pions.  Therefore generically we expect these cumulants to scale as
\beq
\kappa_{m,n}\sim \frac{e^{-(m-n)M\tau}e^{-n m_\pi \tau}}{N^{m+n-1}}\qquad\qquad\qquad (m\ge n)\ .
\eqn{AVcumulants}\eeq
This scaling could be violated if $\Gamma$ can be chosen so that the pion does not couple, as discussed in the calculation of $\kappa_{1,1}$.  Then $m_\pi$ is replaced by $2M$, the mass of a fermion/antifermion pair. Note that at late time, the $1/N$ expansion breaks down in the sense that $\kappa_{2,0}$ becomes smaller than $\kappa_{2,2}$, for example, so long as one is near enough to the chiral limit that $m_\pi < M$. This is despite the fact that $\kappa_{2,2}$ is parametrically smaller by $1/N^2$. However, this breakdown of the $1/N$ expansion will not lead to qualitatively different results because contributions to $\kappa_{m,n}$ which are subleading in $N$ counting will not lead to lighter intermediate states than the leading calculation, unless there is some fortuitous exclusion of the pion at leading order due to the choice of $\Gamma$ that does not persist at higher order.

The above scaling implies that the distribution for the real part of the fermion propagator near the chiral limit becomes highly symmetric about zero at late time because odd moments (for which $(m-n)\ne 0$) are seen to fall off much more quickly than even moments.  This is completely consistent with the Lepage-Savage picture for baryon propagator distributions in QCD.   As in QCD, it will be very difficult to use Monte Carlo methods for the $A/V$ formulation to determine the ground state energy with a fixed large fermion number.  This is not surprising because, like QCD, this theory also has a severe sign problem in the grand canonical formulation for studying states with nonzero fermion density.

%%%%%%%%%%%%%%%%%%%%%
%%%%%%%%%%%%%%%%%%%%%

%\subsubsection{Conclusions about noise in the $A/V$ formulation}

\subsection{Noise distribution in the $\sigma/\pi$ formulation and long-tailed distributions}

\subsubsection{A graphical expansion for cumulants of $Y_\Gamma(\sigma,\pi)$}
\label{TreeGraphs}

We now turn to the task of  computing the cumulants  for the log of the fermion correlator, $Y_\Gamma$ in \eq{corrsp} in the $\sigma/\pi$ formulation.  In this case the cumulants are given by the connected graphs derived from the action
\beq
S_Y = N S[\sigma,\pi]-i s Y_\Gamma[\sigma,\pi]
\eqn{sxsp}\eeq
where $S[\sigma,\pi]$ is given in \eq{ssp}.  The $N$ counting in this formulation is quite different from the $A/V$ case since the $\sigma$ and $\pi$ mesons are singlets under the $U(N)$ symmetry rather than $N\times N$ matrices.  It is convenient to associate the expansion of $ N S[\sigma,\pi]$ in meson fields with vertices labeled by black dots; see Fig.~\ref{fig:blackdots}. There are no black tadpoles, since we have solved for the chiral symmetry breaking vacuum; furthermore there are no explicit black two-point vertices as these are accounted for by using  the full meson propagators.  The expansion of $ i s Y_\Gamma[\sigma,\pi]$ in powers of the meson fields is represented as white dots, the $k$th term  in the expansion drawn as a white vertex connecting  $k$ meson lines; white vertices occur with any number of meson lines, starting with zero, and each is associated with a power of $is$.

The  $n$th cumulant $\kappa_n$ is then given by $n!$ times the sum of connected graphs with $n$ insertions of white vertices, since each white vertex brings a factor of $(is)$ and we have the expansion \eq{cumulantsexp}.  Expanding these graphs in powers of $1/N$ is simple:  each black vertex entails a power of $N$, while each meson propagator  gives a factor of $1/N$.  Loops do not give factors of $N$ since  the $\sigma/\pi$ mesons do not carry $U(N)$ flavor quantum numbers. White dots also do not give factors of $N$.  Thus a contribution to $\kappa_n$ arising from a graph with $n$ white vertices, $b$ black vertices, $p$ meson propagators and $\ell$ loops is of order $N^{b-p}$.  Since every such graph satisfies $p-(b+n) = \ell-1$, we can rewrite the order of the graph as  $N^{-(n-1+\ell)}$.  It follows that the leading contribution to $\kappa_n$ must come from the sum of tree diagrams ($\ell=0$) with $n$ white vertices.   The branches of these connected tree diagrams must end on white tadpoles, and are quite limited in number; the first few leading diagrams are shown in Fig.~\ref{fig:spkappa}.

The sum of tree diagrams can be regarded as the solution to a classical theory, which in the present case is nothing other than the statement that the $1/N$ expansion is equivalent to solving for the generator of cumulants, $\ln \Phi_Y(s)$, as a saddle-point approximation dominated by the classical solution that minimizes the action $S_Y$ (see the Appendix for more details).  Before tackling this calculation it is worthwhile to note several simplifying features of these tree diagrams:

%; a class of subleading contributions to $\kappa_2$ are shown in Fig.~\ref{fig:kappa2sub}.

%%%%%%%%%%%%%%%%%%%%%
\begin{figure}[t]
\includegraphics[width=11 cm]{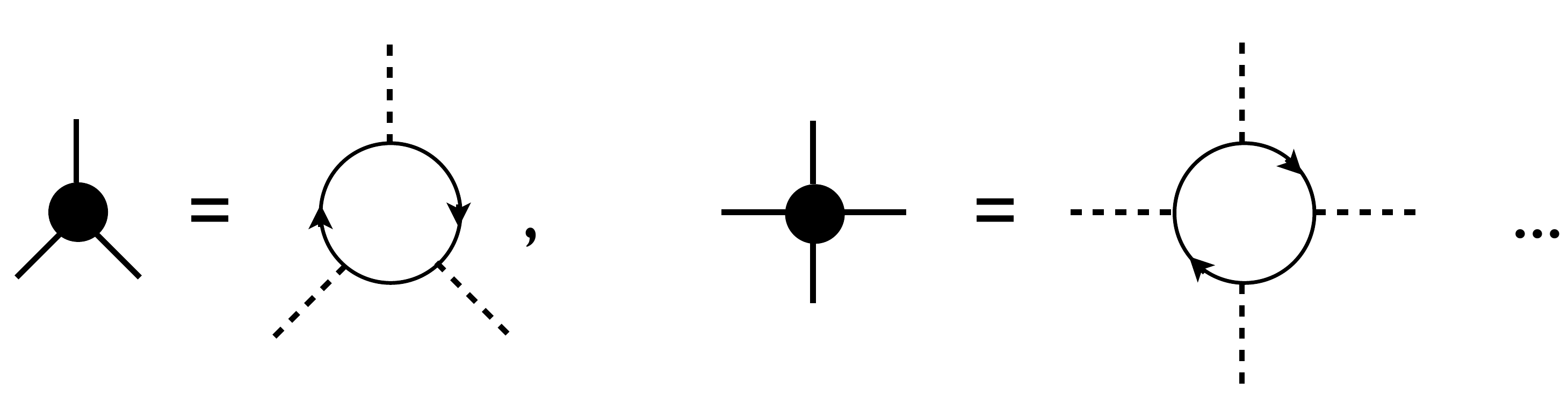}
\caption{\label{fig:blackdots} {\it Black vertices arise in the expansion of the fermion loop, $NS(\sigma,\pi)$ of \eq{ssp} in external meson fields. The tadpole was eliminated by vacuum minimization, and the two-point function gives the exact meson propagators, so these vertices start with the three-point function.}}
\end{figure}
%%%%%%%%%%%%%%%%%%%%%
%%%%%%%%%%%%%%%%%%%%%%
%\begin{figure}[t]
%\includegraphics[width=13 cm]{whitedots.pdf}
%\caption{\label{fig:whitedots} {\it White vertices arise in the expansion of the log of the fermion propagator, $Y_\Gamma$ of \eq{corrsp} in external meson fields. The dotted line indicates that initial and final Dirac indices are contracted with the matrix $\Gamma$.}}
%\end{figure}
%%%%%%%%%%%%%%%%%%%%%%

%%%%%%%%%%%%%%%%%%%%%
\begin{figure}[b]
\includegraphics[width=10cm]{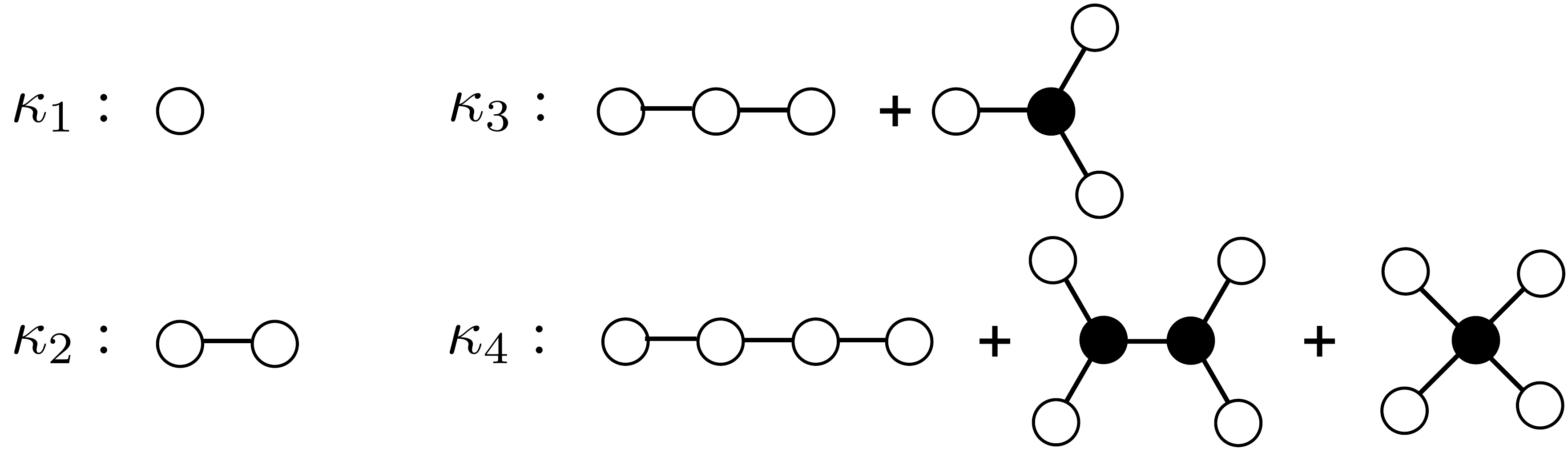}
\caption{\label{fig:spkappa} {\it Leading contributions in the $1/N$ expansion of $\kappa_n$  for $n=1,\ldots,4$ in the $\sigma/\pi$ formulation.  Black vertices are given in Fig.~\ref{fig:blackdots}; white vertices are determined by the expansion of $i s Y[\sigma,\pi]$. Lines represent the exact meson propagators $G_{\sigma,\pi}$ derived from  \eq{spprops}. }}
\end{figure}
%%%%%%%%%%%%%%%%%%%%%

\medskip

\begin{enumerate}[i.]
\item
By choosing a $\Gamma$ that is neither $\gamma_5$ nor $\gamma_\mu\gamma_5$, only the $\sigma$ meson can couple to the white tadpole.  Since parity implies that vertices conserve pion number mod 2, and all tree diagrams end on white tadpole vertices (Fig.~\ref{fig:spkappa}), it follows that the only mesons propagating in these tree graphs are $\sigma$ mesons.   

\item  Since we are defining our observable $Y_\Gamma$ in \eq{corrsp} to be the log of the propagator of a quark with initial and final  2-momentum $\bfp=0$, the meson at the white tadpole vertex must have $\bfp=0$ flowing through it as well.  Then, because all vertices conserve momentum and the leading graphs in Fig.~\ref{fig:spkappa} are all tree diagrams whose branches end on white tadpoles, it follows that $all$ of the internal meson lines in the graph must be at $\bfp=0$, with only nonzero energy flowing through the lines.  

\item 
As we show below, the large $\tau$ behavior of the cumulants arises from graphs with zero energy flowing through the white tadpole; thus, because of conservation of 3-momentum at all vertices, the asymptotic $\tau$ behavior of the cumulants is given by  the graphs with  3-momentum $p$ vanishing everywhere within the graph.

\end{enumerate}
%%%%%%%%%%%%%%%%%%%%%%
%\begin{figure}[b]
%\includegraphics[width=7 cm]{kappa2sub.pdf}
%\caption{\label{fig:kappa2sub} {\it An interesting class of subleading contributions to $\kappa_2$ (two white vertices) at order $N^{-(L+1)}$, where $L$ is the number of loops. In terms of conventional Feynman diagrams, these are ladders with $2,3,4\ldots$ $\sigma/\pi$ meson exchanges between two valence fermion propagators, each of which is traced with the $\Gamma$ matrix. See appendix for explicit examples.}}
%\end{figure}
%%%%%%%%%%%%%%%%%%%%%%

 We now demonstrate the last point (iii): that the white tadpole enforces zero energy flow through the diagram at late time.  To do this we compute the tadpole,  assuming energy $k_1$ and two momentum $\bfk$ flowing out of the meson line:
\beq
\text{tadpole} &=& \frac{-\frac{1}{V}\int_{-\infty}^\infty d\tau' \,\Tr\Gamma \widetilde G (\mathbf{0},\tau'+\tau/2)\delta \sigma(\mathbf{0}, \tau') \widetilde G (\mathbf{0},\tau/2-\tau')}{\frac{1}{V}(2\pi)^2\delta^2(0)\Tr\Gamma \widetilde G (\mathbf{0},\tau/2)} \cr
&=& -\frac{1}{V}\int^{\tau/2}_{-\tau/2} d\tau' \int  \frac{d^3k}{(2\pi)^3} (2\pi)^2\delta^2(\mathbf{k})e^{- i k_1 \tau'}\delta \sigma(k)\cr
&=&  -\frac{1}{V}\int \frac{d^3k}{(2\pi)^3} (2\pi)^2\delta^2(\mathbf{k})\frac{2\, \sin \left( k_1\tau/2\right)}{k_1}\delta \sigma(k)
%\cr &&\cr&
\xrightarrow[\tau\to\infty]{}
%&
-\frac{1}{V}\int  \frac{d^3k}{(2\pi)^3}(2\pi)^3\delta^3(k)\delta \sigma(k)\ ,
\eeq
where $\widetilde G (\mathbf{0},\tau)$ is given in \eq{gtilde0}, provided that 
\beq
z = \Tr\left[ \Gamma\left(\frac{1+\gamma_1}{2}\right)\right] \ne 0\ .
\eqn{disc}\eeq
The $\delta(k_1)$ factor justifies computing the tree graphs in Fig.~\ref{fig:spkappa} with zero 3-momentum flowing through it, as long as we are only interested in the $\tau\to\infty$ behavior of the distribution of propagators. However, note that at finite $\tau$ the factor $\sin ( k_1\tau/2)/k_1$ acts as a filter that still allows $k_1\lesssim 1/\tau$ or, equivalently, which still allows the $\sigma$ field to be time dependent on time scales  $\gtrsim \tau$.  This will be relevant to the next section.

\subsubsection{The generator of cumulants from mean field theory}
\label{MFCalc}

As mentioned previously, the sum of tree graphs is nothing other than the effective action $S_Y=N S- i s Y_\Gamma$ evaluated at its minimum,  a classical solution for the meson fields (see the Appendix for further discussion).  We have shown that for an appropriate choice of $\Gamma$, this solution will in general have a vanishing pion field and a spatially constant, but time dependent, $\sigma(t)$ field.  Finally, we have also shown that the large-$\tau$ behavior is given by an even simpler solution, with a $\sigma$ field that is constant in the two spatial dimensions  and over time $\lesssim \tau$.

It is straightforward to compute $S_Y(\sigma)$ for a constant $\sigma$ field.    Note that $\sigma$ is defined relative to its vacuum value $\vev{\sigma}=(M+m)$, where $M$ is the actual ``constituent" mass of the fermion, while $m$ is the ``current"  mass appearing in the Lagrangian.  For the $N S(\sigma)$ part of $S_Y$ we need only take the expression \eq{Sval} evaluated at $\sigma_S = \sigma- \vev{\sigma}$, up to an overall additive constant:
 \beq
NS(\sigma_S) = -N V  T \left( \frac{3 f\sigma_S^2+6\sigma_S M^2 +2M^3-2\left((\sigma_S+M)^2\right)^{3/2}}{6 \pi}\right)
\eeq
where $V$ is the spatial volume and $ T$ is the time extent of the box. The second part of $S_Y$ is given by 
\beq
i s Y_\Gamma = i s \ln\left[\int\frac{d\omega}{2\pi}\, e^{-i \omega \tau} \Tr\Gamma\frac{1}{-i\omega \gamma_1 + M + \sigma_S}\right] = -is \left((M+\sigma_S)\tau-\ln z\right)\ .
\eeq
The  equation for the minimum of $S_Y$ is therefore given by
\beq
 \left[\sqrt{(M+\sigma_S)^4}-(M^2+f\sigma_S)+\frac{ i \pi s \tau}{N V T}\right]_{\sigma_{S0}} = 0
%\left[ N V  T\, -\sigma_S f -M^2+\sqrt{(\sigma_S+M)^4}+ i \pi s  \tau $\right]_{\sigma_{S0}} = 0
\eeq
with solution
\beq
\sigma_{S0}=\frac{1}{2} \left((f-2 M)+\sqrt{(f-2 M)^2- 4 \pi   i s  \tau/(NV  T)}\right)
\eeq
where we choose the solution that vanishes as $s\to 0$, to recover the correct chiral symmetry breaking vacuum. Plugging this solution back into the effective action, we obtain the generator of cumulants of $Y_\Gamma$, to leading order in the $1/N$ expansion and at late time: 
\beq
\ln\Phi_Y(s) = -S_Y(\sigma_{S0})= is\mu + (2M-f)\tau \, \frac{6 i s\zeta -1 + (1-4 i s \zeta)^{3/2}}{12\zeta}\ ,\cr
\eeq
with
\beq
 \zeta = \frac{\pi \tau}{N V  T (2M-f)^2}\ ,\qquad
 \mu \equiv \ln z-M\tau\ .
\eeq
 Expanding $\ln\Phi_Y(s)$ in powers of $s$ yields the cumulants  $\kappa_n$ for $Y_\Gamma$:
\beq
\kappa_1 =\mu\ ,\qquad
% \kappa_2 =(2M-f) T\zeta\ , \quad
%\kappa_3 = 2(2M-f)T \zeta^2\ , \quad
% \kappa_4 = 12(2M-f)T \zeta^3\ \ldots
\kappa_{n\ge 2} = \frac{(2(n-2))!}{(n-2)!}\, (2 M-f) \tau\, \zeta^{n-1}
\eqn{kappasp} \eeq
 The volume factor in the denominator of  $\zeta $ is easy to understand, arising from the normalization of our one particle states;  the factor of $ T$ is puzzling though, arising from our assumptions of a mean field solution that is constant over the entire spacetime volume.  This does not make sense, since there is no need for $\sigma_S$ to adjust from its chiral symmetry breaking minimum long before or long after the correlator has acted.  As pointed out in the discussion below \eq{disc}, we should not expect $\sigma_S$ to be constant over time scales $\gtrsim \tau$, and in fact we should expect the mean field to relax to its vacuum value for $t\lesssim -\tau/2 $ and $t\gtrsim \tau/2$.  Therefore the factor of $ T$ in $\zeta$  --- the temporal size of the box --- should be replaced by $\sim \tau$, the time scale of the correlator, and we have
 \beq
 \zeta \simeq \frac{\pi }{N V (2M-f)^2}\ \,
\eqn{zeta2} \eeq
 which is independent of $\tau$.  Perhaps one can compute a more accurate, time-dependent mean field solution, but we do not pursue that here.  In an analogous calculation of the Polyakov loop distribution at finite temperature, a strictly time independent mean field solution is probably exact in the large $N$ limit, due to the homogeneity in time of the operator being measured.
%\beq
%\ln\Phi_Y(s) = S_Y[\sigma_0] = \frac{N V_3\left(\sqrt{(f-2 M)^2+\frac{4 i \pi  s T}{N V_3}}+f-2
%   M\right)^2 \left(\sqrt{(f-2 M)^2+\frac{4 i \pi  s T}{N V_3}}-2 f+4
%   M\right)}{24 \pi }-\frac{1}{2} i s \left(2 g+T \left(f+\sqrt{(f-2 M)^2+\frac{4 i \pi
%    s T}{N V_3}}\right)\right)
%\eeq

We conclude this section by remarking that in the particular limit
\beq
N\to\infty\ ,\qquad \tau\to\infty\ ,\qquad \tau \zeta=\text{fixed}
\eeq
all cumulants $\kappa_n$ vanish for $n\ge 3$ and $Y_\Gamma$ assumes a normal distribution, 
\beq
\mathcal{P}(x)=\frac{\sqrt{2\pi}}{\Sigma}e^{\left[-\frac{(x-\mu)^2}{\Sigma^2}\right]}\left(1+\frac{\zeta(x-\mu)}{3\Sigma^4}[(x-\mu)^2-3\Sigma^2]+\mathcal{O}\left(\zeta^2\right)\right) \ ,
\eeq
where $\Sigma^2=(2M-f)\tau \zeta$, which is to say, a Monte Carlo simulation of the fermion propagator will be sampling a log-normal distribution. With $\Sigma^2$ growing linearly with time and a skewness that grows exponentially with $\Sigma$, this distribution will eventually become very heavy-tailed.  Standard simulation methods would fail for such a distribution, but one could use a cumulant expansion of the Monte Carlo data to obtain an accurate measure of the fermion mass, as described in Ref.~\cite{Endres:2011jm}. However, it should be noted that this limit requires $\tau \gtrsim N V$, which is unlikely to be reached in practical lattice simulations of this model. For the more realistic limit $\Sigma^2 \ll 1$ one may use standard statistical techniques and the signal-to-noise ratio in this case will be $\approx 1/\sqrt{N_{\mbox{cfg}}\Sigma^2} \sim \tau^{-1/2}$, where $N_{\mbox{cfg}}$ is the number of gauge field configurations sampled. This power-law dependence on time is far less severe than the exponential falloff of the signal-to-noise ratio in the A/V case [see \Eq{AVcumulants}].

\section{Discussion}
%%%%%%%%%%%%%%%%%%%%%
%\begin{figure*}
%\begin{tabular}{cc}
%\includegraphics[width=7 cm]{fig1aFourBodyLogNormal.pdf}
%\includegraphics[width=7 cm]{fig1bFourBodyNormal.pdf}
%\end{tabular}%
%\caption{\label{fig:ln} {Distribution histograms for $c=C_N(\tau,\phi)$ and $\ln (c)$ for $N=4$ unitary fermions at several times $\tau$, taken from Ref.~\cite{unitary:2011aa}.  Curves fitting $\ln(c)$ are Gaussian, implying that $c$ is approximately log-normal distributed, with $\sigma^2$ increasing with time.  } }
%\end{figure*}
%%%%%%%%%%%%%%%%%%%%%

%%%%%%%%%%%%%%%%%%%%%%
%\begin{figure}[b]
%\includegraphics[width=7 cm]{fig2CumulantDataForDavid.pdf}
%\caption{\label{fig:sigmu} {The quantities $-\frac{1}{E_0}\frac{\partial\mybar y}{\partial\tau}$ and $\frac{1}{E_0}\frac{\partial\sigma^2}{\partial\tau}$ as a function of $N$ for unitary fermions at late times on a lattice of size $L=10$, compared to mean field prediction \eq{mf} (dashed lines).}}
%\end{figure}
%%%%%%%%%%%%%%%%%%%%%%
Our motivation for returning  to the well-worn Nambu-Jona-Lasinio model was to elucidate connections between chiral symmetry breaking and the sign problem in lattice QCD at finite chemical potential, without having to deal with the complications of asymptotic freedom and confinement.  What is particularly attractive about the large-$N$, three-dimensional version of the theory is that (i) it is analytically tractable to compute features of the probability distribution of a fermion correlator, and (ii) the theory has two equivalent formulations, one without a sign problem, and one with a QCD-like, exponentially severe sign problem.

We find that in the QCD-like $A/V$ formulation, the fermion determinant is complex and that a Splittorff-Verbaarschot argument \cite{Splittorff:2006fu, Splittorff:2007ck} can be made to show that the phase of the fermion determinant  has to fluctuate wildly for $\mu>m_\pi/2$, with an expectation value exponentially small in the spatial volume.  When looking at fermion correlators, the distribution evolves to have an exponentially small mean relative to its width, implying a severe signal-to-noise ratio when sampling the correlator using Monte Carlo methods.  Furthermore, the severity of the problem is controlled by the difference between the fermion constituent mass $M$ (playing the role of the baryon mass in QCD) and the much lighter pion mass $m_\pi$.  This follows the Lepage-Savage scaling argument that has even cumulants of the distribution diminishing as a power of $\exp(-m_\pi \tau)$, while the odd cumulants - including the mean - fall off proportional to $\exp(-M\tau)$.  It is interesting that in three dimensions one can choose an observable for measuring the fermion mass [by a particular choice of the matrix $\Gamma$ in \eq{corrAV}] which eliminates the coupling of the fermion/antifermion pair to the pion, and thereby eliminates the problem of noise in the measurement of the fermion mass.  Such a trick might be a useful way to reduce the noise in simulations of QCD in four dimensions, even if it cannot eliminate it.

In contrast, the $\sigma/\pi$ formulation with even $N$ has no sign problem at nonzero $\mu$, and the correlator distribution - the cumulants of which can be analytically computed - is, in a certain limit, log-normal.  For extremely long times the distribution is heavy-tailed, which can pose challenges to Monte Carlo sampling, but this sort of problem seems to be  less severe than the exponential falloff of the signal-to-noise ratio in the $A/V$ formulation as seen with the cumulant expansion analysis of Refs.~\cite{Endres:2011jm,Endres:2011er, Endres:2012cw}. For more moderate times, standard statistical methods should apply, and the signal-to-noise ratio in this case is found to have only power-law suppression with time. It should be noted that such distributions have been seen in QCD for intermediate times, before any asymptotic pion noise sets in. It has been hypothesized that these distributions are related to elastic scattering between particles \cite{DeGrand:2012ik}; the volume factors in our expressions for the cumulants, \eqstwo{kappasp}{zeta2}, give support to this picture.

Our analysis should make it clear that the sign problem encountered in QCD at nonzero chemical potential is not a fermion problem, but instead a consequence of interactions.  In particular, if the particles being studied can exist in a tightly bound state of valence fermions, there is going to be a sign problem - a generic feature of a theory with dynamical chiral symmetry breaking, in which a light composite pion emerges as a Goldstone boson.  This is what happens in the $A/V$ formulation of the NJL model studied here: the fact that the $A$ and $V$ fields will bind a fermion/antifermion pair into a light or massless pion implies that studies of the fermion correlator will be noisy, and that at nonzero chemical potential for the fermion there will be a sign problem.  In the $\sigma/\pi$ formulation without a sign problem, the pion exists as a fundamental field and not as a bound state.

The lessons learned from this  model raise the question: is it possible to introduce a mean field for the pion into our formulation of lattice QCD (without changing the theory) so that the pion does not appear as a  bound state of a valence quark/anti-quark pair, $ \bar Q Q$?  For example, one might add and subtract a four-fermion interaction to the QCD Lagrangian; the attractive one could be introduced by $\sigma$ and $\pi$ auxiliary fields, while the repulsive interaction could be derived by means of auxiliary $A$ and $V$ fields.  Then a valence $\bar Q Q$ pair would feel the usual gluon attraction, but $A/V$ repulsion, and so would not bind to form a light pion.  Instead, pions would appear as fundamental fields that could be created through $\bar Q Q$ annihilation, but would not appear as bound states of valence quarks. It is expected that the conventional sign problem could be ameliorated in such a theory - but this example probably introduces other sign problems, a cure perhaps as devastating as the disease.  

Nevertheless, we believe that inventing a way to introduce the pions into QCD as fundamental fields could be an important step toward solving the QCD sign problem and beginning to study the properties of ordinary and dense matter from first principles.

\appendix
\section{Connection Between Mean Field Calculation and Tree Level Cumulant Diagrams}
In Sec. \ref{MFCalc}, we asserted that the sum of tree graphs contributing to the cumulants of $Y_{\Gamma}$ is equal to the effective action $S_Y$, evaluated at its minimum. Here, we will show this equivalence in more detail.

We have two different representations for the generating function, $W(s)$. The first is in terms of a functional integral:
\beq
Z = e^{-W(s)} = \int [d\phi] \,e^{-N S(\phi) + i s Y(\phi)}\ .
\eeq
The second representation is as the generating function for cumulants:
\beq
W(s) =\text{const} - \sum_{n=1}^\infty \frac{(i s)^n}{n!} \kappa_n\ .
\eeq
Changing variables, $s= N r$, we have
\beq
e^{-W(r)} = \int [d\phi] \,e^{-N \CA (r,\phi) }\ ,
\eeq
with $ \CA(r,\phi) = S(\phi) + i r Y(\phi)$, and  
\beq
W(r) =\text{const} - N\sum_{n=1}^\infty \frac{(i r)^n}{n!} \, \left(N^{n-1}\kappa_n\right)\ .
\eeq
We  now compute $W$ in a large-$N$ expansion, which is equivalent to a mean field expansion:
\beq
\CA(r,\phi)  = \CA^{(0)}+ \sum_{n=2}^\infty \frac{\CA^{(n)}}{n!} \delta\phi^n\ ,
\eeq
where we have defined
\beq
\CA^{(n)} = \frac{\delta^n\CA}{\delta\phi^n}\Biggl\vert_{\phi=\phi_0}\ ,\qquad \CA^{(1)}=0\ ,
\eeq
with $\phi_0$ being the classical solution that minimizes $\CA$ and $\delta\phi = (\phi-\phi_0)$. This allows us to write
\beq
e^{-W(r)} &=& e^{-N\CA^{(0)}}\int [d\delta\phi]\, e^{-N\sum_{n=2}^\infty \frac{\CA^{(n)}}{n!} \delta\phi^n}\cr&&\cr
 &=&
 e^{-N\CA^{(0)}}\left[e^{-N \sum_{n=3}^\infty \frac{\CA^{(n)}}{n!} \frac{\delta^n}{\delta J^n}}\int [d\delta\phi] \, e^{-\frac{N}{2}\CA^{(2)}\delta\phi^2 + J\delta\phi}\right]_{J=0}\cr&&\cr
 &=& \frac{ \text{const}}{\sqrt{\det N \CA^{(2)} } }\times e^{-N\CA^{(0)}}\times\left[e^{-N \sum_{n=3}^\infty \frac{\CA^{(n)}}{n!} \frac{\delta^n}{\delta J^n}} \, e^{-\frac{J^2}{N \CA^{(2)}}}\right]_{J=0} \ .
\eqn{master}\eeq
To determine the leading contributions to $\kappa_n$, we must locate the terms that are both leading in $N$ and $r$-dependent. 
\begin{enumerate}[i]
\item
The constant factor in \eq{master}  is independent of $r$ and does not contribute to $\kappa_n$.  
\item
The factor in brackets  in \eq{master} is the sum of connected diagrams whose propagators scale as $1/N$ and vertices as $N$. Thus, these diagrams scale as
\beq
N^{V-P} = N^{1-L}\ ,
\eeq
using the topological invariant $L+V-P =1$, where $V, P,$ and $L$ are the numbers of vertices, propagators, and loops respectively. These diagrams do not contain tadpoles, as there are no terms linear in $\phi$. Therefore, this quantity only contributes to $W(r)$ at one loop and higher and is thus $\mathcal{O}(N^0)$ and subleading.  
\item
The determinant factor in \eq{master} contributes to $W(r)$ a term $\half \Tr\ln N \CA^{(2)} = \text{const} + \half\Tr\ln \CA^{(2)}$. The constant is $N$ dependent, but $r$ independent, whereas the second term is $r$ dependent, but $N$ independent.  Therefore, the determinant factor comes in at $\mathcal{O}(N^0)$ and is also subleading. 
\item
We are left with the $e^{-N\CA^{(0)}}$ term: this  gives $N \CA^{0}$, the classical action at its minimum, as the leading contribution to $W$, at $\mathcal{O}(N)$.
\end{enumerate}

We will now relate $\CA^{(0)}$, the action at the classical minimum, to the diagrams in Fig.~\ref{fig:spkappa}. From above, we see that we can expand $W(r)$ in powers of $N$:
\beq
W(r) &=&\text{const} +  N \left[W_0(r) + N^{-1} W_1(r) + N^{-2} W_{2}(r) + \ldots \right]\ ,
\eeq
where $W_0(r) = N \CA^{(0)}$. We can also write $W(r)$ in terms of cumulants,
\beq
W(r)=  \text{const} - N\sum_{n=1}^\infty \frac{(i r)^n}{n!} \, \left(N^{n-1}\kappa_n\right)\ .
\eeq
From this, we see that $\kappa_n$ may be expanded as
\beq
\kappa_n = \frac{1}{N^{n-1}} \sum_{p=0}^\infty \frac{k_{n,p}}{N^p} \ ,
\eeq
with
\beq
W_p(r) = -\sum_{n=1}^\infty \frac{(i r)^n}{n!} \, k_{n,p} \ .
\eeq
We have previously identified the sum of tree graphs with $n$ white vertices as the leading contribution to the $n$th cumulant, $k_{n, 0}$. Thus, we see explicitly that
\beq
W_0 = -\sum_{n=1}^\infty \frac{(i r)^n}{n!} \, k_{n,0}\ ,
\eeq
which proves our assertion that the sum of tree graphs in Fig.~\ref{fig:spkappa} is equal to the effective action $S_Y=N S(\phi) - i s Y(\phi)$, evaluated at its minimum.

\begin{acknowledgments}
We would like to acknowledge stimulating conversations with Thomas DeGrand, Michael Endres, Jong-Wan Lee, Anyi Li, and Martin Savage. This work was supported in part by U.S.\ DOE Grants No.\ DE-FG02-00ER41132 and No.\ DE-FG02-93ER-40762.
\end{acknowledgments}
\include{MFPaper}
\bibliography{GNref10}
\end{document}